# A zero-one SUBEXP-dimension law for BPP

Philippe Moser *

**Abstract**

We show that BPP has either SUBEXP-dimension zero (randomness is easy) or BPP = EXP (randomness is intractable).

## 1 Introduction

Lutz's resource bounded-measure [6] allows to quantify the size of standard complexity classes such as BPP, NP, ... where any class can be either small, large or in between, corresponding to having $p$-measure $0, 1$ or being non-measurable ($p$ stands for polynomial-time computable martingales). It was recently refined by the introduction of effective dimension [7] with (among others) the notion of $p$-dimension. One of the motivations is to quantify the structure of sets that have resource-bounded measure zero, in the same way as classical Hausdorff dimension can be used to quantify many sets of Lebesgue measure zero. Examples include classes of languages decidable by (nonuniform) circuits that all have measure zero, but whose dimension differs depending on a parameter of the circuit size (see [3]).

Among many applications, resource-bounded measure was used in derandomization, when van Melkebeek showed that the standard probabilistic complexity classes ZPP and BPP have either $p$-measure zero or one [11]. Impagliazzo and Moser later proved that RP also satisfies the zero-one law [4]. These results ruled out the non-measurable case, thus yielding further evidence that if randomness is not intractable (i.e. $C \neq$ EXP, where $C \in \{$ZPP, RP, BPP$\}$) then randomness is easy (i.e. $C$ has $p$-measure zero).

In this paper we improve van Melkebeek's zero-one law for BPP [11], by strengthening the statement "randomness is easy" in two aspects; first by reducing the time-bounds of the martingales from polynomial to subpolynomial and second by replacing measure with dimension.

Informally, a class has $p$-measure zero, if there is a polynomial time computable predictor (a martingale) that can predict infinitely often any language $L$ in the class without making too many mistakes. Quantifying how often a martingale makes correct predictions about a language $L$, can be measured by a real number $s \in [0, 1]$, where $s = 0$ means that the martingale predicts $L$ correctly on a large fraction of all strings, whereas $s = 1$ means the predictions are correct on a small fraction. The $p$-dimension of a $p$-measure zero class $C$, is the infimum over all $s \in [0, 1]$ for which there is a polynomial time martingale that predicts all languages in $C$ correctly on at least a fraction $s$ of all strings.

Lutz's formulation of resource bounded-dimension [7] (similarly to resource-bounded measure) is a general theory where the time bound of the martingales is a parameter that can be set depending on what classes one tries to measure in. For example choosing polynomial time bounds yields the notion of $p$-dimension which is a dimension notion on the class E.

---

*Department of Computer Science, National University of Ireland, Maynooth Co. Kildare, Ireland. Email: pmoser(at)cs.nuim.ie

Although it is easy to extend the theory by considering time bounds greater than polynomial (i.e. measuring in classes containing E), it is not easily extended to smaller time bounds (i.e. subpolynomial bounds that allow to measure in classes like P or SUBEXP for example). It is only recently that a dimension notion on subclasses of E was obtained in [9]. Previous measure notions on subclasses of E were introduced prior to [9] (see [1, 10] for measures on P (and above), and [8] for a measure on PSPACE), but they cannot be used to define dimension because the associated martingales cannot place enough bets (see [9] for more details).

In this paper we use the dimension on SUBEXP from [9], corresponding to subpolynomial time computable martingales (instead of polynomial ones for $p$-dimension on E), to show that either BPP has dimension zero or BPP = EXP. Our proof uses van Melkebeek's argument [11] in the dimension on SUBEXP setting introduced in [9]. Because subpolynomial martingales are weaker than polynomial ones, our result immediately holds for Lutz's $p$-dimension.

This result seems to be particular to BPP, and it is not obvious whether it carries over to ZPP or RP.

## 2 Preliminaries

For $\epsilon > 0$, denote by $\mathsf{E}_\epsilon$ the class $\mathsf{E}_\epsilon = \bigcup_{\delta < \epsilon} \mathsf{DTIME}(2^{n^\delta})$. SUBEXP is the class $\cap_{\epsilon > 0} \mathsf{E}_\epsilon$. Let us fix some notations for strings and languages. Let $s_0, s_1, \ldots$ be the standard enumeration of the strings in $\{0, 1\}^*$ in lexicographical order, where $s_0 = \lambda$ denotes the empty string. Denote by $s_0^n, s_1^n, \cdots, s_{2^n-1}^n$ all strings of size $n$ ordered lexicographically, i.e. $s_t^n = s_{2^n + t - 1}$ for any $0 \leq t \leq 2^n - 1$. A sequence is an element of $\{0,1\}^\infty$. If $w$ is a string or a sequence and $0 \leq i < |w|$ then $w[i]$ and $w[s_i]$ denotes the $i$th bit of $w$. Similarly $w[i \ldots j]$ and $w[s_i \ldots s_j]$ denote the $i$th through $j$th bits. We identify language $L$ with its characteristic function $\chi_L$, where $\chi_L$ is the sequence such that $\chi_L[i] = 1$ iff $s_i \in L$. $L \upharpoonright s_n$ stands for $L[s_0 \cdots s_n]$. If $w_1$ is a string and $w_2$ is a string or a sequence extending $w_1$, we write $w_1 \sqsubseteq w_2$.

The Shannon entropy is the following continuous function

$$H : [0,1] \to [0,1]$$
$$H(\alpha) = \alpha \log \frac{1}{\alpha} + (1 - \alpha) \log \frac{1}{1 - \alpha}$$

where $H(0) = H(1) := 0$.

### 2.1 SUBEXP-dimension

Lutz's resource-bounded measure [6] is obtained by imposing appropriate resource bounds on a game-theoretical characterization of classical Lebesgue measure, in which a gambler places bets on the successive membership bits of a hidden language $A$. The game proceeds in infinitely many rounds where at the end of round $n$, it is revealed to the gambler whether $s_n \in A$ or not. The game starts with capital 1. Then, in round $n$, depending on the first $n-1$ outcomes $w = \chi_A[0 \ldots n-1]$, the gambler bets a certain fraction $a \in [0,1]$ (resp. $1-a$) of his current capital, that the $n$th word $s_n \in A$ (resp. $s_n \notin A$). The game is fair, i.e. the amount placed on the correct event is doubled, the one placed on the wrong guess is lost. The player wins on a language $A$ if he manages to make his capital arbitrarily large during the game. The wins of the player are measured by functions called rate-martingales.

**Definition 2.1** *[9] A rate-martingale is a function $D : \{0,1\}^* \to [0,2]$ such that for every $w \in \{0,1\}^*$, $D(w0) + D(w1) = 2$.*

A rate-martingale outputs the factor by which the capital is increased after the bet.

SUBEXP-dimension [9] is defined via rate-martingale families which share their wins. These rate-martingales are computed by Turing machines with random access to their input, i.e. machines that have oracle access to their input and can query any bit of it. To enable such machines to compute the length of their input without reading it, they are given $s_{|w|}$ together with input $w$; this convention is denoted by $M^w(s_{|w|})$.

**Definition 2.2** *[9] Let $\alpha > 0$. An $\mathsf{E}_\alpha$-family of rate-martingales $(\{D_i\}_i, \{Q_i\}_i, \mathrm{ind})$, is a family of rate-martingales $\{D_i\}_i$, where*

- *$Q_i : \mathbb{N} \to \mathcal{P}(\{0,1\}^*)$ are disjoint $\mathsf{E}_\alpha$-printable query sets (i.e. there is a Turing machine that on input $(i, 1^n)$ outputs all strings in $Q_i(n)$ in $2^{n^\alpha} + i^c$ steps with $c \geq 1$), i.e. $Q_i(n) \cap Q_j(n) = \emptyset$ (for every $i \neq j$), $\cup_{i \leq \frac{2^n}{n}} Q_i(n) = \{0,1\}^{\leq n}$ and $Q_i(m) \subseteq Q_i(n)$ for $m < n$,*

- *$\mathrm{ind} : \{0,1\}^* \to \mathbb{N}$ is a $\mathsf{DTIME}(2^{n^\alpha})$-computable function, such that $D_i(L \upharpoonright x)$ is computable by a random access Turing machine $M$ in time $2^{|x|^\alpha}$ i.e. $M^{L \upharpoonright x}(x, i) = D_i(L \upharpoonright x)$ where $M$ queries its oracle only on strings in $Q_i(|x|)$, and $\mathrm{ind}(x)$ is the index such that $x \in Q_{\mathrm{ind}(x)}(|x|)$.*

Each rate-martingale $D_i$ of the family only bets on strings inside its query set $Q_i$ (i.e. $D_i(x) = 1$ for $x \notin Q_i(|x|)$). The function ind on input a string $x$, outputs which rate-martingale is to (possibly) bet on $x$. The idea is that the rate-martingales share their wins, and have the ability to divide the bets along all members of the family. We are interested in the total capital such a family wins.

**Definition 2.3** *[9] The win function of an $\mathsf{E}_\alpha$-family of rate-martingales $(\{D_i\}_i, \{Q_i\}_i, \mathrm{ind})$ is the function $W_{\{D_i\}_i} : \{0,1\}^* \to \mathbb{Q}$, where $W_{\{D_i\}_i}(L \upharpoonright x) = \prod_{i \leq \frac{2^{|x|}}{|x|}} \prod_{y \leq x} D_i(L \upharpoonright y)$.*

Because $D_i(L \upharpoonright x)$ is the factor by which the capital is multiplied after the bet on $x$, the product in Definition 2.3 is exactly the total capital the whole family of rate-martingales would win, would they be able to share their wins after each bet.

Lutz's key idea to define resource-bounded dimension on $\mathsf{E}$ [7] is to tax the martingales' wins. The following definition formalizes this tax rate notion (the definition of $s$-success for standard martingales is from [2], in [9] the definition was extended to families of rate martingales).

**Definition 2.4** *[9] Let $s \in [0,1]$, $(\{D_i\}_i, \{Q_i\}_i, \mathrm{ind})$ be an $\mathsf{E}_\alpha$-family of rate-martingales, and let $L$ be a language. We say $\{D_i\}_i$ $s$-succeeds on $L$, if $\limsup_{n \to \infty} 2^{(s-1)n} W_{\{D_i\}_i}(L \upharpoonright s_{n-1}) = \infty$.*

Similarly $D$ $s$-succeeds on class $C$, if $D$ $s$-succeeds on every language in $C$.

The SUBEXP-dimension of a complexity class is the highest tax rate that can be levied on the martingales' wins without preventing them from succeeding on the class.

**Definition 2.5** *[9] Let $C$ be a class of languages. The $\mathsf{E}_\alpha$-dimension of $C$ is defined as $\dim_{\mathsf{E}_\alpha}(C) = \inf\{s \in [0, \infty) : \text{ there is an } \mathsf{E}_\alpha\text{-family of rate-martingales } \{D_i\}_i \text{ that } s\text{-succeeds on } C\}$.*

$\mathsf{E}_\alpha$-dimension satisfies the following union property.

**Theorem 2.1** *[9] Let $0 < \delta < \alpha$, $\{C_j\}_j$ be a family of classes, and let $\{t_j\}_j$ with $t_j \in [0,1]$ such that for every $\epsilon > 0$ there exists an $\mathsf{E}_\delta$-family of rate-martingales $\{D_{i,j}\}_{i,j}$ with identical query sets such that $\{D_{i,j}\}_i$ $(t_j + \epsilon)$-succeeds on $C_j$. Then for $C = \bigcup_j C_j$, $\dim_{\mathsf{E}_\alpha}(C) \leq \sup_j\{t_j\}$.*

## 2.2 Derandomization of BPP

We need the following special case of a derandomization result for BPP under a uniform hypothesis from [5].

**Theorem 2.2** *[5] If $\mathsf{BPP} \neq \mathsf{EXP}$ then for every language $A \in \mathsf{BPP}$ and every $\delta > 0$, there exists a language $B \in \mathsf{DTIME}(2^{n^\delta})$, such that for infinitely many lengths $m$*

$$\Pr_{|x|=m}[A(x) = B(x)] > 1 - \frac{1}{m}. \tag{1}$$

# 3 BPP satisfies the zero-one law for dimension in SUBEXP

The following result shows with respect to SUBEXP gamblers, randomness is either weak or intractable.

**Theorem 3.1** *If $\mathsf{BPP} \neq \mathsf{EXP}$ then for every $\alpha > 0$, BPP has $\mathsf{E}_\alpha$-dimension 0.*

*Proof.* We need the following lemma.

**Lemma 3.1** *Let $C$ be closed under polynomial many-one reductions and such that for every $\delta > 0$ and for every language $A \in C$ there exists a language $B \in \mathsf{DTIME}(2^{n^\delta})$, such that for infinitely many lengths $m$*

$$\Pr_{|x|=m}[A(x) = B(x)] > 1 - \frac{1}{m}. \tag{2}$$

*Then for every $\alpha > 0$, $C$ has $\mathsf{E}_\alpha$-dimension 0.*

Let us prove the lemma. Let $\alpha > 0$, $\epsilon > 0$. Let us show that $\dim_{\mathsf{E}_\alpha}(C) < \epsilon$. Let $\delta < \alpha' < \alpha$ and $\epsilon' < \epsilon$ (to be determined later).

Let $B \in \mathsf{DTIME}(2^{\lfloor n^\delta \rfloor})$. Consider the following family of rate-martingales $(\{D_{B,j}\}_{j\in\mathbb{N}}, \{Q_j\}_{j\in\mathbb{N}}, \mathrm{ind})$ where the query sets are obtained by partitioning $\{0,1\}^k$ into $2^{k-\lfloor k^\delta \rfloor}$ query sets, i.e. let

$$T_j^k := \{s_{j2^{\lfloor k^\delta \rfloor}}^k, \ldots, s_{(j+1)2^{\lfloor k^\delta \rfloor}-1}^k\} \quad \text{and} \quad Q_j(n) = \cup_{k=1}^n T_j^k \quad \text{(for all integers } n\text{)}$$

with the convention that $Q_j(n)$ is empty when $j \geq 2^{n-\lfloor n^\delta \rfloor}$. The corresponding index function ind is obvious.

$D_{B,j}$ on $T_j^n$ bets according to the following list of "strategies" $\{p_i\}_{i\in\mathbb{Z}_+}$, where $p_i$ only bets on the strings $s_{\lfloor \epsilon' 2^i \rfloor - 1}, \ldots, s_{2^i-2}$ and bets that the corresponding membership bits are given by the sequence $B[s^i_{\lfloor \epsilon' 2^i \rfloor}, \ldots, s^i_{2^i-1}]$. Thus on $T_j^n$ the only $p_i$'s that bet are the ones for which

$$z_n := \log(2^n + j2^{\lfloor n^\delta \rfloor} + 2) \leq i \leq -\log(\epsilon') + \log(2^n - 2 + (j+1)2^{\lfloor n^\delta \rfloor}) =: z'_n.$$

$D_{B,j}$ upon starting betting on the first string in $T_j^n$, divides its current capital $c$ ($c$ is unknown but will cancel out in the final equation) into shares $\{c2^{-i}\}_{i \in \mathbb{Z}_+}$ where $c2^{-i}$ is devoted to $p_i$. Hence after betting on the first string, the new capital $c'$ is given by

$$c' = \sum_{i \geq 1} c 2^{-i} D'_i(L \upharpoonright s^n_{j2\lfloor n^\delta \rfloor})$$

$$= c[\sum_{i=1}^{z_n-1} 2^{-i} + \sum_{i \geq z'_n+1} 2^{-i} + \sum_{i=z_n}^{z'_n} 2^{-i} D'_i(L \upharpoonright s^n_{j2\lfloor n^\delta \rfloor})]$$

$$= c[1 - \sum_{i=z_n}^{z'_n} 2^{-i} + \sum_{i=z_n}^{z'_n} 2^{-i} D'_i(L \upharpoonright s^n_{j2\lfloor n^\delta \rfloor})]$$

where

$$D'_i(L \upharpoonright s_t) = \begin{cases} 2(1-2\epsilon') & \text{if } \lfloor \epsilon' 2^i \rfloor - 1 \leq t \leq 2^i - 2 \text{ and } L(s_t) = B(s^i_{t+1}) \\ 4\epsilon' & \text{if } \lfloor \epsilon' 2^i \rfloor - 1 \leq t \leq 2^i - 2 \text{ and } L(s_t) \neq B(s^i_{t+1}) \\ 1 & \text{otherwise (no bet).} \end{cases}$$

Thus

$$D_{B,j}(L \upharpoonright s^n_{j2\lfloor n^\delta \rfloor}) = \frac{c'}{c} = 1 - \sum_{i=z_n}^{z'_n} 2^{-i} + \sum_{i=z_n}^{z'_n} 2^{-i} D'_i(L \upharpoonright s^n_{j2\lfloor n^\delta \rfloor})$$

and for any $s_t \in T_j^n$

$$D_{B,j}(L \upharpoonright s_t) = \frac{1 - \sum_{i=z_n}^{z'_n} 2^{-i} + \sum_{i=z_n}^{z'_n} 2^{-i} \prod_{y \in T_j^n \cap \{s_0, \ldots, s_t\}} D'_i(L \upharpoonright y)}{1 - \sum_{i=z_n}^{z'_n} 2^{-i} + \sum_{i=z_n}^{z'_n} 2^{-i} \prod_{y \in T_j^n \cap \{s_0, \ldots, s_{t-1}\}} D'_i(L \upharpoonright y)}.$$

Since both $z'_n < 2n$ (for $n$ big enough), and $D'_i(L \upharpoonright s_t)$ (with $s_t \in T_j^n$) requires computing $B$ on strings of size at most $2n$, $D_{B,j}(L \upharpoonright s_t)$ can be computed in time less than $an2^{(2n)^\delta}2^{n^\delta}$ (for some constant $a$), which is smaller than $2^{n^{\alpha'}}$ by an appropriate choice of $\delta$. Thus $(\{D_{B,j}\}_{j \in \mathbb{N}}, \{Q_j\}_{j \in \mathbb{N}}, \text{ind})$ is a $2^{n^{\alpha'}}$-family of rate-martingales.

Let us show that this family $\epsilon'$-succeeds on $C$. Let $A \in C$ and let $D$ be the language where for every length $k$

$$D[s_1^k \ldots s_{2^k-1}^k] = A[s_0 \ldots s_{2^k-2}]$$

and let $B \in \mathsf{DTIME}(2^{\lfloor n^\delta \rfloor})$ be the language predicting $D$ on infinitely many length $i$, i.e. there exists an infinite set $M$ such that for every $i \in M$

$$|\{x : |x| = i \text{ and } B(x) = D(x)\}| \geq (1 - \frac{1}{i})2^i.$$

Let $m \in M$. By construction of $\{D_{B,j}\}_j$, the $m$-th strategy $D'_m$ of $\{D_{B,j}\}$ bets correctly on at least $2^m(1 - 2\epsilon')$ bits of $A[s_{\lfloor \epsilon' 2^m \rfloor - 1} \ldots s_{2^m - 2}]$ (for $m$ big enough), therefore because

$Q_j(m-1) = \emptyset$ whenever $j > 2^{m-1}2^{-\lfloor m-1 \rfloor^\delta} - 1 =: v_m$ we have

$$W_{\{D_{B,j}\}_j}(A \upharpoonright s_{2^m-2}) = \prod_{j=1}^{v_m} \prod_{y \leq s_{2^m-2}} D_{B,j}(A \upharpoonright y)$$

$$= \prod_{j=1}^{v_m} \prod_{k=1}^{m-1} \prod_{y \in T_j^k} D_{B,j}(A \upharpoonright y)$$

$$= \prod_{j=1}^{v_m} \prod_{k=1}^{m-1} [1 - \sum_{i=z_k}^{z'_k} 2^{-i} + \sum_{i=z_k}^{z'_k} 2^{-i} \prod_{y \in T_j^k} D'_i(L \upharpoonright y)]$$

$$\geq \prod_{j=1}^{v_m} \prod_{k=1}^{m-1} [2^{-m} \prod_{y \in T_j^k} D'_m(L \upharpoonright y)]$$

$$= \prod_{j=1}^{v_m} 2^{-m(m-1)} \prod_{y \in Q_j(m-1)} D'_m(L \upharpoonright y)$$

$$\geq 2^{-m(m-1)v_m} [2(1-2\epsilon')]^{(1-2\epsilon')2^m} [4\epsilon']^{(2\epsilon')2^m}$$

$$= 2^{-m(m-1)v_m} 2^{(1-H(2\epsilon'))2^m}.$$

This implies that $\{D_{B,j}\}_j$ $\epsilon$-succeeds on $A$, because for every $m \in M$

$$\frac{W_{\{D_{B,j}\}_j}(A \upharpoonright s_{2^m-2})}{2^{(1-\epsilon)(2^m-3)}} \geq 2^{-m(m-1)v_m + [\epsilon - H(2\epsilon')]2^m}$$

$$\geq 2^{[-\frac{m(m-1)}{2^{\lfloor (m-1)^\delta \rfloor +1}} + \epsilon - H(2\epsilon')]2^m}$$

which grows unbounded by choosing $\epsilon'$ small enough such that $-\frac{m(m-1)}{2^{\lfloor (m-1)^\delta \rfloor +1}} + \epsilon - H(2\epsilon') > 0$ (for $m$ large enough).

Let $\{B_i\}_i$ be an enumeration of $\mathsf{DTIME}(2^{n^\delta})$, and let $D_{i,j} := D_{B_i,j}$ be the rate-martingale defined as above. By construction, the query sets do not depend on language $B_i$, hence $\{D_{i,j}\}_{i,j}$ is an $\mathsf{E}_{\alpha'}$-family of rate-martingales, such that for every $\epsilon > 0$, $\{D_{i,j}\}_j$ $\epsilon$-succeeds on $X_i = \{A \in C | \exists^\infty m : \Pr_{|x|=m}[A(x) = B_i(x)] > 1 - \frac{1}{m}\}$. By Theorem 2.1, $X = \cup_i X_i$ has $\mathsf{E}_\alpha$-dimension 0 (by choosing $\alpha' < \alpha$ accordingly). Since $C \subseteq X$, this ends the proof of Lemma 3.1. Since Theorem 2.2 implies the hypothesis of Lemma 3.1 with $C = \mathsf{BPP}$, the proof is complete. □

## Final Remark

It is not known whether a similar result holds for $\mathsf{ZPP}$ or $\mathsf{RP}$. The martingales used in the proof of the zero-one law for $\mathsf{RP}$ from [4] do not bet often enough to yield a dimension result. It would be interesting to see whether any dimension type result can be obtained for these two classes.

## Acknowledgment

I thank John Hitchcock for commenting an earlier draft of this paper, and pointing out an improvement from dimension 1/2 to 0.